\documentclass[a4paper,fleqn,usenatbib]{mnras}

\usepackage{newtxtext,newtxmath}

\usepackage[T1]{fontenc}
\usepackage{ae,aecompl}

\usepackage{graphicx}	
\usepackage{amsmath}	
\usepackage{amssymb}	

\title[Kinetic theory of the strahl]{Kinetic theory of the electron strahl in the solar wind}

\author[Stanislav Boldyrev et al.]{
Stanislav Boldyrev,$^{1, 2}$\thanks{E-mail: boldyrev@wisc.edu}
Konstantinos Horaites$^{1}$
\\
$^{1}$Department of Physics, University of Wisconsin -- Madison, 1150 University Avenue, Madison, WI 53706, USA\\
$^{2}$Space Science Institute, Boulder, CO 80301, USA\\
$^{3}$Another Department, Different Institution, Street Address, City Postal Code, Country
}

\date{Submitted April 27, 2019}

\pubyear{2015}

\begin{document}
\label{firstpage}
\pagerange{\pageref{firstpage}--\pageref{lastpage}}
\maketitle

\begin{abstract}
We develop a kinetic theory for {the electron strahl, a beam of energetic electrons which propagate} from the sun along the Parker-spiral-shaped magnetic field lines. By assuming a Maxwellian electron distribution function in the near-sun region where the plasma is collisional, we derive the strahl distribution function at larger heliospheric distances. We consider the two most important mechanisms that broaden the strahl: Coulomb collisions and interactions with oblique ambient whistler turbulence (anomalous diffusion).  We propose that the energy regimes where these mechanisms are important are separated by {an approximate threshold}, ${\cal E}_c$; for the electron kinetic energies ${\cal E}<{\cal E}_c$ the strahl width is {mostly} governed by Coulomb collisions, while for ${\cal E}>{\cal E}_c$ by interactions with the whistlers. The Coulomb broadening decreases as the electron energy increases; the whistler-dominated broadening, on the contrary, increases with energy and it can lead to efficient isotropization of energetic electrons and to formation of the electron halo.  The threshold energy ${\cal E}_c$ is relatively high in the regions closer to the sun, and it gradually decreases with the distance, implying that the anomalous diffusion becomes progressively more important at large heliospheric distances. At $1$~AU, {we estimate} the energy threshold to be about ${\cal E}_c\sim 200\,{\rm eV}$.      
\end{abstract}




\section{Introduction}
The solar wind consists of a magnetized plasma nearly radially propagating from the sun. Observations show that the temperature of the expanding plasma declines with the radial distance $r$, following an approximate power law trend $T(r)\propto r^{-\gamma}$, where $\gamma\approx 0.5$ \cite[e.g.,][]{koehnlein96,cranmer2009,stverak15,bale2016,chen2016,verscharen2019}. In particular, at the distance of $1$~AU the solar wind cools down to about $10~\mbox{eV}$ as compared to the solar corona where the plasma temperature is on the order of $100$~eV. The solar wind plasma is, however, weakly collisional, so while the temperature of its Maxwellian core follows the mentioned trend rather well, the velocity distribution function also has features that deviate significantly from the thermal Maxwellian distribution. In particular, the electron velocity distribution function (eVDF) can be represented as consisting of three major components, the nearly Maxwellian thermal core, the suprathermal beam aligned with the direction of the magnetic field (the so-called strahl), and the nearly isotropic and broad (non-Maxwellian) halo, which overlaps in energy with the strahl \cite[e.g.,][]{feldman75,pilipp87,pierrard2016}. The strahl and halo are relatively tenuous, for instance, at $1$~AU they comprise about $5\%$ of the total electron density. However, since their energies exceed that of the core by an order of magnitude, the heat flux associated with the strahl is non-negligible and it can heat the solar wind at relatively large heliospheric distances \cite[e.g.,][]{stverak09,stverak15}. Moreover, non-Maxwellian anisotropic distribution function can be a source of kinetic instabilities and small-scale turbulence that lead to formation of structures, particle heating and acceleration \cite[e.g., ][]{forslund70}. 

{In this work we develop a kinetic theory of the electron strahl. {In order to understand how the strahl is formed and how it changes with the radial distance, we trace the evolution of the electron velocity distribution function all the way from the hot inner region ($\sim 5 - 10\, R_\odot$) where the electrons are assumed to have a given distribution (say, a Maxwellian) to larger heliospheric distances.} {We assume that the magnetic field has a Parker-spiraled structure, and solve the drift-kinetic equation that describes the evolution of the electron distribution function along the magnetic field lines.} {The energetic, nearly collisionless electrons streaming along the magnetic-field lines away from the sun, get collimated into a narrow beam (strahl) due to conservation of their magnetic moment. Weak Coulomb collisions with the background plasma, on the other hand, tend to broaden their collimation angle.}
 
{Comparison with some strahl measurements in the fast solar wind \cite[e.g.,][]{stverak09,stverak15,horaites18a,horaites18b} demonstrates that our Coulomb theory allows one to describe the physics of strahl formation rather well, on both qualitative and quantitative levels. In particular, the theory is able to predict the strahl angular broadening and the number of particles in the strahl.} {When the Coulomb theory provides a good agreement with the observational data, it means that other possible scattering mechanisms, such as interactions of the electrons with plasma fluctuations, are relatively unimportant \cite[][]{ogilvie00,pierrard01,horaites15,horaites18a,horaites18b,horaites2019}}.   

{In some measurements, however, the angular distribution of the strahl electrons is wider than the distribution predicted by the Coulomb model \cite[e.g.,][]{anderson12,graham17,graham18}. In such cases, it is reasonable to assume that the enhanced broadening is provided by ambient plasma turbulence that scatter the energetic electrons (anomalous scattering); such scattering should be taken into account in addition to that provided by Coulomb collisions (classical scattering).} One of the natural candidates for the anomalous scattering is whistler turbulence \cite[e.g.,][]{gary75, gary94,vocks03,vocks05,saitogary07,pagel07,pierrard2011,wilson13,lacombe14,kajdic16,stansby16,tang2018}. In order to analyze the strahl broadening caused by whistler turbulence, we assume that the turbulence is oblique with respect to the background magnetic field, and  incorporate the anomalous scattering in the drift kinetic equation. In this respect our consideration is complementary to previous studies that considered electron-strahl broadening caused by the whistlers propagating {\em along} the background magnetic field \cite[][]{pierrard2011,tang2018}.  

{We find that the scattering by the whistlers rapidly increases with the intensity of the turbulence. As a result, the two scattering mechanisms (that is, Coulomb scattering and scattering by the oblique whistlers) dominate in different regions of the phase space roughly separated by an energy threshold, ${\cal E}_c$. The strahl electrons with lower energies, ${\cal E} < {\cal E}_c$, are mostly scattered by classical Coulomb collisions, while the more energetic electrons, with energies exceeding ${\cal E}_c$, by whistler turbulence. As the anomalous scattering becomes far more significant than Coulomb collisions at high energies, it may lead to isotropization of the energetic strahl electrons and to formation of the electron halo. We estimate that at $1$~AU, the threshold energy may be on order of $200$~eV. The dominance of Coulomb collisions at lower energies and the predicted energy-dependent strahl broadening at higher energies is qualitatively consistent with the recent analytical and observational studies \cite[e.g.,][]{horaites2019,bercic2019}.}

\section{The Coulomb theory of the strahl}
In this section, we develop a kinetic theory for the strahl component of the eVDF taking into account classical Coulomb collisions and neglecting anomalous scattering effects.  The speed of the strahl electrons is significantly larger than that of the solar wind. The  suprathermal electrons not only experience significantly weaker Coulomb collisions as compared to the core electrons, but they also stream from the sun to very large distances ($\sim 10~\mbox{AU}$) along nearly stationary magnetic-field lines. Indeed, the magnetic field lines are advected with the speed of the solar wind, while the speed of the electrons is much higher. 

When the collision frequency is much smaller than the gyrofrequency of the particles, the electron velocity  distribution function is gyrotropic; it can be averaged over the fast period or electron gyromotion.  It can then be written using the variables of velocity $v$, the {cosine of the angle between velocity and the (anti-sunward directed) background magnetic field} $\mu\equiv \cos\theta=v_\|/v$, and the distance along a magnetic field line $x$. The distribution obeys the following drift-kinetic equation \cite[e.g.,][]{kulsrud2005,horaites15}:
\begin{eqnarray}
\label{drift_kinetic}
\frac{\partial f}{\partial t}+\mu v \frac{\partial f}{\partial x} - \frac{1}{2}\frac{d\ln B}{dx}v \left(1-\mu^2\right) \frac{\partial f}{\partial \mu} \nonumber \\
 - \frac{e E_\parallel}{m_e} \left[\frac{1-\mu^2}{v} \frac{\partial f}{\partial \mu}  + \mu \frac{\partial f}{\partial v}\right] 
 = \hat C(f).
\end{eqnarray}
In this equation, {$E_\|=-\nabla \phi(x) \cdot{\hat x}$} is the electric field along the magnetic field line, and ${\hat C}(f)$ denotes the collision integral. 

Let us first consider a purely collisionless evolution{, $\hat C(f) = 0$}, and assume that we are interested in a steady-state distribution.  {Equation~(\ref{drift_kinetic})} then takes an especially simple form if one uses the following variables: the magnetic moment $M=m_ev_\perp^2/(2B(x))$, the total energy $E=m_ev^2/2+e\phi(x)$, and the distance $x$. As can be directly verified, the electron velocity distribution function $f(E, M, x)$ then obeys the equation
\begin{eqnarray}\label{drift_kinetic_short}
\mu v \,{\partial f}/{\partial x}=0,
\end{eqnarray}
meaning that the distribution function is independent of the distance. The magnetic field lines that generally follow the Parker-spiral configuration, are almost radial close to the sun.  By using the observationally inferred trends for the temperature and density variations with the heliospheric distance \cite[e.g.,][]{koehnlein96,stverak15}, one can expect that at a distance of approximately $x= r_0\sim 5 - 10\, R_\odot$, the plasma is dense and relatively more collisional than at larger radial distances, so one may assume that the electron distribution is Maxwellian with a temperature of about $T_0\sim 100~{\rm eV}$. (This simplifying assumption, although plausible, is not essential for our kinetic derivation. Our theory can be generalized for any gyrotropic distribution specified at distance $r_0$. Our goal is to find the electron distribution function at larger distances, $r>r_0$, once this inner-region distribution is known.) 

In the new variables, the Maxwellian distribution looks like 
\begin{eqnarray}\label{f_maxwellian}
f(r_0,E,M)=A_0\exp\left\{-\frac{E}{T_{0}} \right\}\theta\left(E-M B_0\right),
\end{eqnarray}
where $A_0=n_0(m_e/2\pi T_{0})^{3/2}$ is the normalization coefficient, $B_0=B(r_0)$, and the theta function reflects the fact that our variables obey the restriction $E\geq MB_0$.  According to Eq.~(\ref{drift_kinetic_short}), the distribution function in these variables is independent of the distance, $f(r, E, M)=f(r_0, E, M)$; we therefore obtain from Eq.~(\ref{f_maxwellian}), $r>r_0$:
\begin{eqnarray}\label{f_beam}
f(r,E,M)=A_0\exp\left\{-\frac{m_ev^2}{2T_{0}}-\frac{e\phi(r)}{T_{0}} \right\}\theta\left(v^2+\frac{2e}{m_e}\phi(r)-\frac{B_0}{B(r)}v_\perp^2 \right).
\end{eqnarray} 
Away from the sun, $r> r_0$, the ambipolar potential energy $e\phi(r)$ fast approaches its maximal value, $e\phi_\infty$, which is  a few times larger than~$T_0$.\footnote{We define the ambipolar potential in such a way that it is zero at $r=r_0$. Using standard methods, one can demonstrate that it increases to the values comparable to its asymptotic value $\phi_\infty$ at a typical distance of order $r\gtrsim r_0(m_e/m_i)^{1/4}$, which for a hydrogen plasma gives $r\gtrsim 6\,r_0$.} {We can estimate the pitch-angle breadth ($\theta$) of the beam at such distances, by equating the argument of the theta-function in Eq.~(\ref{f_beam_2}) to zero:}
\begin{eqnarray}\label{f_beam_2}
\sin^2 \theta=\frac{v_\perp^2}{v^2}= \frac{B(r)}{B_0}\left(1+\frac{2e\phi_\infty}{m_e v^2}\right).
\end{eqnarray}
For instance, at $r =1~{\rm AU}$, we can estimate by order of magnitude that ${B(r)}/{B_0}\sim 10^{-4}$, so for the electron kinetic energy of $m_ev^2/2\sim 100~{\rm eV}$, the strahl collimation angle would be rather narrow, $\theta\sim 10^{-2}$. 

One can show, however, that such a narrow collimation angle cannot be established, since it will be broadened by weak Coulomb collisions. In order to describe the Coulomb collisions, we need to add the collision integral in Eq.~(\ref{drift_kinetic_short}). The energetic strahl electrons have relatively weak energy exchange with the plasma particles forming the Maxwellian core, however, they experience a significant pitch-angle scattering. In order to describe the strahl broadening, we, therefore, retain in the collision integral only the term describing the pitch-angle scattering \cite[e.g.,][]{helandersigmar02}:
\begin{eqnarray}\label{coll_op_eq}
\hat C(f)  = \left(\frac{4 \pi n(x) e^4 \Lambda \beta }{m_e^2 v^3}\right) \frac{\partial}{\partial \mu} \left(1-\mu^2\right)\frac{\partial f}{\partial \mu}.
\end{eqnarray}  
This collision integral describes the pitch-angle scattering of the suprathermal strahl electrons $(v^2\gg v^2_{Te})$ by the Maxwellian core electrons and the core ions. In Eq.~(\ref{coll_op_eq}), $\beta=(1+Z_{eff})/2$, where $Z_{eff}$ is the effective ion charge. For the solar wind plasma, $\beta$ can be estimated as $\beta\approx 1.05$.  The Coulomb logarithm can be estimated at $1~{\rm AU}$ as $\Lambda\approx 30$, it is a slowly varying function of the distance, and $n(r)$ is the density of the core electrons, which is approximately equal to the density of the ions, see e.g.,~\cite{horaites2019}. 

The collision integral can be rewritten using the new variables $E$, $M$, and $x$, which gives for the steady-state drift-kinetic equation \cite[][]{horaites2019}:
\begin{eqnarray}
\frac{\partial f(x, E, M)}{\partial x} =\frac{4 \pi e^4 \Lambda \beta  n(x)}{{\cal E}(E,x)B(x)}\frac{\partial }{\partial M}M\sqrt{1-\frac{MB(x)}{{\cal E}(E,x)}} \frac{\partial f}{\partial M},\label{diffusion}
\end{eqnarray}
where we have denoted ${\cal E} (E, x)\equiv E-e\phi(x)=m_ev^2/2$. The expression in the square root in Eq.~(\ref{diffusion}) can be simplified since, as one can directly verify, ${MB(x)}/{{\cal E}(E,x)}=v_\perp^2/v^2=\sin^2\theta \ll 1$, and this term can be neglected. Moreover, as we are interested in the runaway electrons, the total electron energy should exceed the ambipolar potential barrier, $E> e\phi_\infty$. 

{An equation similar to Eq.~(\ref{diffusion}) was analyzed in our previous treatment of the problem \cite[][]{horaites2019}, where we were interested in the evolution of the strahl at radial distances significantly exceeding the coronal region, $r\gg r_0$. As a result, we were able to obtain the angular distribution of the strahl electrons, but could not specify the electron distribution function uniquely -- our solution contained an arbitrary function of the electron kinetic energy. In the present work, we relate the electron distribution function to the boundary condition at $r\sim r_0$, and derive a complete solution for the suprathermal electron strahl.}

It is convenient to represent the kinetic energy in the form ${\cal E}= \Delta E+\mathcal{T}(x)$, where the first term, $\Delta E= E-e\phi_\infty$, is independent of the distance, and all the radial dependence is included in the function $\mathcal{T}(x)=e\phi_\infty-e\phi(x)$.\footnote{{We remind the reader that we denote by~$x$ the distance along a magnetic field line, while keep the variable~$r$ for the radial distance. One variable can be expressed through the other by using the Parker-spiral shape of the magnetic field.}} {One can argue that the function ${\cal T}(x)$ is on the order of the local electron temperature, ${\cal T}(x)\approx T(x)$. Indeed, the local kinetic energy of an electron, $E-e\phi(x)$, cannot exceed $e\phi_\infty-e\phi(x)$, otherwise, such an electron will run away to infinity. Therefore, ${\cal T}(x)$ is on the order of the local kinetic energy of the background electrons, and, therefore, is proportional to their temperature. In what follows, we, therefore, will approximate ${\cal E}\approx  \Delta E+T(x)$.}   

The form of Equation~(\ref{diffusion}) suggests that we may introduce the following new variable (see also \cite[][]{horaites2019}):
\begin{eqnarray}
dy=\left(\frac{4\pi e^4\Lambda \beta}{{\cal E}} \right)\left(\frac{n(x)}{B(x)} \right)\,dx.\label{dy}
\end{eqnarray}
In this equation, $dx$ is the length element along a magnetic field line.  Since the magnetic field is frozen into the solar-wind flow, the combination $dx n(x)/B(x)$ is an invariant of the motion. We can, therefore, evaluate it at the distance $r_0$ close to the sun, where the magnetic field lines are nearly radial, $dx n(x)/B(x)=dr_0 n(r_0)/B(r_0)$. Next, we notice that the solar wind speed is nearly constant as a function of the radial distance at $r>r_0$. Therefore, any two points separated by the radial distance $dr$ that corresponds to the separation $dx$ along a field line, do not change their radial separation during the motion, thus $dr=dr_0$.  We can, therefore, write
\begin{eqnarray}
dy=\left(\frac{4\pi e^4\Lambda \beta}{{\cal E}} \right)\left(\frac{n_0}{B_0} \right)\,dr =\left(\frac{4\pi e^4\Lambda \beta}{\Delta E+\mathcal{T}(r)} \right)\left(\frac{n_0}{B_0} \right)\,dr.\label{dy_dr}
\end{eqnarray}

This equation can, in principle, be integrated once the temperature profile is specified. For instance, if we assume the model power-law behavior for the electron temperature $\mathcal{T}(r)\propto r^{-0.5}$, a straightforward calculation will give
\begin{eqnarray}\label{y_r}
y=\frac{4\pi e^4 \Lambda \beta n_0}{B_0\Delta E}\,R(r),
\end{eqnarray}
where we have denoted
\begin{eqnarray}\label{R_r}
R(r)=r\left[1-2\left(\frac{\mathcal{T}(r)}{\Delta E}\right) +2\left(\frac{\mathcal{T}(r)}{\Delta E}\right)^{2}\log\left(\frac{\Delta E}{\mathcal{T}(r)}+1\right)\right].
\end{eqnarray}
For a given energy, $R(r)$ is a function of the heliospheric distance only. 
Since we are interested in suprathermal electrons, we can often approximate ${\cal E}\approx \Delta E\gg \mathcal{T}(r)$, and, therefore, in this limit the function $R(r)$ can be replaced by $r$ in Eq.~(\ref{y_r}). We, however, note that in cases when we need to evaluate exponential functions, we may need to keep the small ${\cal T}(r)$ term in the general expression for the energy $E=e\phi_\infty +\Delta E=e\phi_\infty+{\cal E}-{\cal T}(r)$.

Equation~(\ref{diffusion}) now turns into a two-dimensional diffusion equation in $M$-space:
\begin{eqnarray}\label{diffusion_2D}
\frac{\partial}{\partial y}f{(y,E,M)}=\frac{\partial}{\partial M}M\frac{\partial}{\partial M}f.
\end{eqnarray}
The standard solution of Eq.~(\ref{diffusion}) takes the form \citep[Eqs. 5-6, ][]{horaites2019}:
\begin{eqnarray}\label{diffusion_2D_gen_sol}
f(y,E,M)=\frac{C(E)}{y} \exp \Big( -\frac{M}{y}\Big),
\end{eqnarray} 
where $C(E)$ is an arbitrary function. 
{ We would now like to match the solution (\ref{diffusion_2D_gen_sol}) with our formula for the collisionless case (\ref{f_beam}). We will accomplish this by imposing that the two solutions have the same width ($\Delta M$) and amplitude at some distance $y_m$. First let us consider that the width of the strahl (in terms of $M$) inferred from Eq.~(\ref{diffusion_2D_gen_sol}), $\Delta M$, is on the order $\Delta M \sim y$. By comparison, the width of the strahl described by Eq.~(\ref{f_beam}) is estimated as $\Delta M\sim E/B_0$. So we find that the solutions will have the same width, and can be approximately matched, at the (energy-dependent) distance $y_m \sim E / B_0$. Additionally equating the amplitudes of solutions (\ref{diffusion_2D_gen_sol}), (\ref{f_beam}) leads to the expression, which we use to model the strahl\footnote{From Eq.~(\ref{y_r}), one can estimate that the distance at which the two solutions match is several times larger than $r_0$, so that the collimation angle of the suprathermal electrons is smaller than one, and the diffusion equation~(\ref{diffusion_2D}) derived in the limit of small collimation angles, is applicable.} at distances $y\gtrsim y_m$: 
\begin{eqnarray}\label{diff_sol}
f=A_0\exp\left(-\frac{E}{T_0}\right)\frac{E}{B_0}\frac{1}{y}\exp\left(-\frac{M}{y}\right).
\end{eqnarray}
The obtained solution can be re-written in a more compact form if we introduce the electron mean free path at~$r=r_0$, defined as
\begin{eqnarray}
\lambda_0=\frac{T_0^2}{4\pi n_0 e^4\Lambda \beta}.
\end{eqnarray}  
We then get for the suprathermal part of the distribution function
\begin{eqnarray}\label{f_total}
f\approx A_0 F_0 \frac{\lambda_0}{R(r)}\left[\frac{{\Delta E}+e\phi_\infty}{e\phi_\infty}\right]\frac{\Delta E}{T_0}\exp\left(-\frac{\Delta E}{T_0}\right)\times \nonumber \\
\times \exp\left(-\frac{{\cal E}{\Delta E}\sin^2\theta}{T_0^2}\frac{\lambda_0}{R(r)}\frac{B_0}{B(r)} \right),
\end{eqnarray}
where
\begin{eqnarray}\label{F_0}
F_0=\frac{e\phi_\infty}{T_0}\exp\left(-\frac{e\phi_\infty}{T_0} \right)\approx \left(\frac{T_{i,0}}{T_0}\right)^{1/2}\left(\frac{m_e}{m_i} \right)^{1/2}.
\end{eqnarray}
We remind the reader that  ${\cal E}=m_ev^2/2$ is the electron kinetic energy, $\Delta E\approx {\cal E}-{T}(r)$ is the excess of the kinetic energy of the strahl electrons over the thermal energy of the background plasma, and the distance parameter $R(r)$ is given by formula~(\ref{R_r}). In formula (\ref{F_0}), $T_{i, 0}$ is the temperature of the ions at the source location~$r_0$. The estimate for $F_0$ comes from the fact that the ambipolar potential barrier $\phi_\infty$ is established as to ensure that the proton and electron fluxes from the sun balance each other. The electrons, as lighter particles, escape with higher velocities, therefore, $\phi_\infty$ is negative and $e\phi_\infty$ is positive. Observations and analytical modeling suggests that the ions are heated more efficiently in the corona, so that $T_{i, 0}/T_0\approx 10$ \cite[e.g.,][]{chandran11}. A kinetic calculation, assuming that at $r_0$ the distributions of both the ions and the electrons are Maxwellian and the outflows are radially symmetric, then leads  to the estimate~(\ref{F_0}) and to $e\phi_\infty/T_0\approx 4$.\footnote{Strictly speaking, the kinetic calculation gives the following condition for the potential barrier $\left[1+\frac{e\phi_\infty}{T_0}\right]\exp\left(-\frac{e\phi_\infty}{T_0} \right)= \left(\frac{T_{i,0}}{T_0}\right)^{1/2}\left(\frac{m_e}{m_i} \right)^{1/2}$. However, as $e\phi_\infty/T_0\approx 4$, we may neglect unity in the square brackets and use Eq.~(\ref{F_0}) as an estimate.}  Formula~(\ref{f_total}) is the main result of our Coulomb theory of the electron strahl. 

\section{{Analysis of the Coulomb strahl solution}}
In this section we discuss what predictions follow from the strahl solution mediated by Coulomb collisions (Eq.~\ref{f_total}) and  to what extend they agree with the available observations.   

First result is the width of the strahl, which can be found from the exponential factor in Eq.~(\ref{f_total}):
\begin{eqnarray}\label{sin}
\sin^2\theta\approx \frac{T_0^2}{{\cal E}\Delta E}\,\frac{R(r)}{\lambda_0}\,\frac{B(r)}{B_0}.
\end{eqnarray}
The formula is valid as long as our main assumption $\sin^2\theta\ll 1$ is satisfied. In the limit ${\cal E}\gg {\cal T}(r)$, this formula is consistent with the result derived previously in \cite{horaites2019}, where it was found to be in good agreement with the Wind measurements using the SWE instrument \cite[][]{ogilvie00}. We remind  that the Parker-spiral magnetic-field strength has the form
\begin{eqnarray}
B(r)=B_0\,\frac{r_0^2}{r^2}\,\sqrt{1+\frac{r^2}{r^2_{45}}},\label{B}
\end{eqnarray} 
where $r_{45}$ is the heliospheric distance where the magnetic field line makes an angle of $45^\circ$ with the radial direction. From observations, one can estimate that $r_{45}\approx 1~{\rm AU}$. From Eqs.~(\ref{sin}) and (\ref{B}) one can see that the strahl becomes progressively stronger collimated with the distance in the inner heliosphere, $r< r_{45}$. However, at $r\gg r_{45}$ the width of the strahl saturates, that is, it becomes independent of the distance. This effect was discovered in \cite[][]{horaites2019}. It can be explained in the following way. At large heliospheric distances, the Parker spiral becomes progressively better aligned with the azimuthal heliospheric direction, so that the travel distance of the electrons increases as they propagate away from the sun, which enhances the efficiency of the Coulomb collisions. Simultaneously, the strength of the magnetic field~(\ref{B}) declines with the distance slower in the outer heliosphere, which reduces the magnetic focusing effect. The solution presented above demonstrates that in this case the magnetic focusing and Coulomb pitch-angle broadening balance each other at $r\gg r_{45}$, which leads to a universal saturated width of the strahl.   

Second, formula~(\ref{f_total}) also allows us to estimate the number of particles in the strahl. First, we note that due to the exponential cutoff, only the energies $\Delta E\approx {\cal E}\lesssim T_0$ will contribute significantly to the integral of the distribution function~(\ref{f_total}). Therefore, the expression in the square brackets in Eq.~(\ref{f_total}) is of order unity. Next, we assume that the strahl is narrow, so we can approximate $\sin\theta\sim\theta$. The  strahl distribution function~(\ref{f_total}) can then be easily integrated over the velocity space, and we obtain 
\begin{eqnarray}\label{n_st}
\frac{n_{st}(r)}{n(r)}\approx \frac{F_0}{2}\frac{B(r)}{B_0}\frac{n_0}{n(r)}\exp\left(\frac{{\cal T}(r)}{T_0}\right).
\end{eqnarray}
We remind that ${\cal T}(r)$ is on the order of the local electron temperature, and ${\cal T}(r)$ is smaller than $T_0$. At $1$~AU, we estimate from this formula that $n_{st}(r)/n(r)\approx 0.05$. This simple derivation   provides a rather good agreement with the values inferred from observations \cite[e.g.,][]{stverak09}. Due to the slowly changing function $\exp\left({{\cal T}(r)}/{T_0}\right)$, our formula (\ref{n_st}) also predicts that in the inner heliosphere, the fraction of particles in the strahl slowly declines with the distance, which is in a agreement with the observations by~\cite{stverak09}.  

In the outer heliosphere $r\gg r_{45}$, however, our Coulomb formula predicts a relative increase of the strahl fraction, while the observations demonstrate the opposite trend.  This may be not surprising, however, since our Coulomb model does not include possible strong angular scattering and isotropization of the strahl electrons due to non-Coulomb effects, and, therefore, it overestimates the strahl population. Non-Coulomb (anomalous) broadening may also explain the instances where the strahl width was observed to be broader than that predicted by the Coulomb model or where the width of the strahl was found to increase with the heliospheric distance rather than decrease or saturate \cite[e.g.,][]{anderson12,graham17,graham18,horaites2019}. The non-Coulomb scattering effects are discussed in section~\ref{anomalous}. 

Third, as follows from Eq.~(\ref{sin}), the strahl width is independent of the parameters of the source -- the electron temperature $T_0$ and the magnetic field $B_0$. The information about the electron distribution function of the source is, however, imprinted in the strahl amplitude. Our formula~(\ref{f_total}) demonstrates that for the Maxwellian velocity distribution of the source electrons, the strahl amplitude is proportional to $({\cal E}/T_0)\exp(-{\cal E}/T_0)$. This result agrees with the exponential fall-off of the strahl amplitude previously reported in the SWE measurements by \cite{ogilvie00}, where the characteristic temperature scales of about $T_0\sim 100$~eV were detected. We also note that the strahl amplitude, as given by our formula~(\ref{f_total}), is rather low. An estimate shows that at $1$~AU , the strahl component of the distribution function starts to exceed the Maxwellian core component at about ${\cal E}\gtrsim 4\, T(r) $, which also agrees with available observations~\cite[e.g.,][]{stverak09}. 

Finally, it is interesting to point out that a non-monotone velocity profile of the strahl, as given by the Coulomb theory (\ref{f_total}), may, in principle, lead to an instability, and if so, it would hardly persist at large heliospheric distances. If the core$+$strahl distribution function  becomes unstable, it will quickly relax to a stable monotone profile. The relaxation process will smooth out the velocity profile at energies $\Delta E\lesssim T_0$, but will not change the number of particles in the strahl as estimated in Eq.~(\ref{n_st}), and the exponential decline of the strahl amplitude at higher energies, $\Delta E \gtrsim T_0$.

\section{Anomalous broadening of the strahl}
\label{anomalous}
Observations demonstrate that the electron strahl overlaps in energies with another suprathermal  component of the electron distribution function, the so-called halo. The halo population is nearly isotropic in the velocity space, and its distribution is well approximated by a power-law function at large energies \cite[e.g.,][]{pierrard2016}. The origin of the halo is currently not well understood. It is possible that several distinct mechanisms are at play in the halo formation. One mechanism is related to the possibility that the fast electrons can be trapped by the magnetic field lines at large heliospheric distances and directed back toward the sun by reflection by  plasma inhomogeneities  or by following looped magnetic field lines \cite[e.g.,][]{scudderolbert79,gosling1993,gosling2001,horaites2019}.  Indeed, the observed isotropy of the halo demonstrates the presence of sun-ward moving energetic electrons. Since these electron are rather energetic, they are virtually unaffected by Coulomb collisions and, therefore, they can come from very large radial distances ($\sim 10$~AU). As these electrons propagate closer to the sun in the regions of increasing magnetic-field strength,  magnetic de-focusing can efficiently isotropize their velocity distribution function. The halo electrons can thus be the population of suprathermal electrons escaped the sun as a strahl but later trapped by magnetic field lines at global heliospheric scales, and isotropized by the combination of Coulomb collisions and magnetic de-focusing. This is consistent with the fact that the halo is nearly isotropic but the strahls are predominantly observed in the anti-sunward directions. 

An alternative possibility, which will be discussed in more detail below, is that the halo is generated locally from the strahl electrons that experience very strong angular scattering by some mechanism \cite[e.g.,][]{stverak09,stverak15}. The nature of such a mechanism can be debated, but a possible candidate for scattering is interaction with ambient plasma turbulence, in particular, the whistler modes. The wave-particle resonance condition, $\omega-k_\| v -n\Omega_e=0$, can be easily satisfied for $n=\pm 1$.  The quantitative description of this process depends on the model assumed for the whistler turbulence. For instance, one can assume that turbulence consists mostly of the whistlers propagating along the direction of the magnetic field lines, $k_\|\gg k_\perp$. Such models were developed in \cite[e.g.,][]{pierrard2011,tang2018}, {as possible candidates for explaining the evolution of suprathermal electrons.  The advection-diffusion kinetic equations describing the electron strahl were derived that could be analyzed analytically and numerically.}  

In our consideration, we concentrate on the complementary case, when the whistler turbulence is oblique, that is, $k_\perp > k_\|$. This assumption may be consistent with some phenomenological and numerical models \cite[e.g.,][]{cho2009,boldyrev_p12,meyrand2013} and observations \cite[][]{alexandrova09,kiyani09a,chen10b,chen12a,sahraoui13a,narita16}, and similarly to the case of quasi-parallel turbulence, it also allows for analytical treatment.  In the case of oblique propagation, the whistler-mode frequency has the form
\begin{eqnarray}\label{omega}
\omega=k_\|k_\perp v_A d_i,
\end{eqnarray} 
where $v_A$ is the Alfv\'en speed and $d_i$ is the ion inertial scale. Whistlers exist in the region of the phase space $\omega \gg kv_{Ti}$, where $v_{Ti}$~is the thermal velocity of the ions. In the case when the ion plasma beta is of order one, $\beta_i=v_{Ti}^2/v_A^2\approx 1$, this condition implies that $k_\|d_i\gg 1$. For electron velocities $v\gg v_{Te}$, we see from Eq.~(\ref{omega}) that $\omega \ll k_\|v$, therefore, the resonance condition reads $k_\|v=\pm\Omega_e$. 

The simplest analytical description of the wave-particle interaction is in the form of quasilinear diffusion, which demonstrates how the distribution function evolves under the action of a large number of particle interactions with an ensemble of linear waves \cite[e.g.,][Chapter 17]{stix1992}.  This is certainly an approximation as the whistler modes are not necessarily linear waves. However, it is known from analytic modeling and observation, that even in the case of strongly nonlinear turbulence, the linear and nonlinear terms in the governing plasma equations are on the same order (the so-called critical balance condition) \cite[][]{goldreich_toward_1995,cho2009,tenbarge2012,boldyrev_p12}. Therefore, a consideration based on a linear dispersion relation, in addition to being analytically tractable, provides a good order-of-magnitude estimate. One can ask what contribution the quasilinear interaction provides to the pitch-angle scattering. For that we write the collision operator as 
\begin{eqnarray}\label{C1}
{\hat C}=S\frac{\partial}{\partial \mu} \left(1-\mu^2\right)\frac{\partial f}{\partial \mu},
\end{eqnarray} 
where $S=S_{C}+S_{QL}$ is the sum of the Coulomb collision term and the quasilinear diffusion term.
The Coulomb collision part is given by Eq.~(\ref{coll_op_eq}). The quasilinear diffusion coefficient $S_{QL}$ is proportional to the integral of the intensity of the electric-field fluctuations associated with the whistler waves \cite[][page 498]{stix1992}. In our case of $\omega\ll k_\|v=\Omega_e$, this coefficient takes the form
\begin{eqnarray}\label{SQL}
S_{QL}=\frac{\pi e^2 \Omega_e^2}{m_e^2 v^3} \int  d^2 k_\perp \frac{1}{\omega^2}\left\vert E(k_\|, k_\perp)\right\vert^2_{k_\|=\Omega_e/v}.
\end{eqnarray}
The electric field of oblique whistler modes has a strong potential component, which is related to their magnetic component as $(\beta_e/2) e\phi_k/T_e\sim \delta B_k/B$, where $B$ is a constant background magnetic field \cite[e.g.,][]{chen_boldyrev2017}. This allows us to  express the electric spectrum through the magnetic spectrum,
\begin{eqnarray}\label{E}
\left\vert E(k_\|, k_\perp)\right\vert^2=k_\perp^2\frac{4 T_e^2}{e^2 \beta^2_e}\left\vert\frac{\delta B_k}{B}\right\vert^2.
\end{eqnarray}
Substituting this result and expression (\ref{omega}) for the whistler frequency, in the integral~(\ref{SQL}), we obtain
\begin{eqnarray}\label{SQL_2}
S_{QL}=\frac{4\pi \Omega_e^2T_e^2}{m_e^2v^3\beta_e^2v_A^2d_i^2k_\|^2}\int d^2 k_\perp\, \left\vert \frac{\delta B_k}{B}\right\vert^2_{k_\|=\Omega_e/v}.
\end{eqnarray}
Conveniently, the scattering coefficient provided by oblique whistler modes depends only on the field-parallel spectrum of the magnetic fluctuations, for which we will assume a power-law behavior
\begin{eqnarray}\label{B_par}
\int d^2 k_\perp\, \left\vert \frac{\delta B_k}{B}\right\vert^2=\left\vert \frac{\delta B_{k_\|}}{B}\right\vert^2=Dk_\|^{-\alpha}.
\end{eqnarray}
Here $D$ is the normalization coefficient. It is convenient to express this coefficient through the intensity of magnetic fluctuations in whistler turbulence. Since whistlers exist only at scales $k_\|d_i\gg 1$ (and they are strongly Landau damped at $k_\|d_i\approx 1$ \cite[e.g.,][]{chen_b2013}), we estimate the total magnetic energy in the whistler fluctuations as
\begin{eqnarray}\label{norm}
\left(\frac{\delta B}{B}\right)^2=\int\limits_{1/d_i} dk_\| \left\vert \frac{\delta B_{k_\|}}{B}\right\vert^2=\frac{d_i^{\alpha-1}}{\alpha-1} D,
\end{eqnarray}
which gives
\begin{eqnarray}\label{D}
D= \frac{\alpha-1}{d_i^{\alpha-1}} \left(\frac{\delta B}{B}\right)^2.
\end{eqnarray}
The intensity of the whistler magnetic fluctuations is a parameter of the theory; it can be inferred, for example, from observations, or obtained from numerical simulations or analytical modeling.  Substituting expression~(\ref{D}) into Eq.~(\ref{B_par}) and into the quasilinear diffusion integral~(\ref{SQL_2}) we finally arrive at the expression for the scattering coefficient
\begin{eqnarray}\label{coll_op_eq_2}
S  = \frac{4 \pi n(r) e^4 \Lambda  }{m_e^2 v^3}\left[1+\frac{4\pi (\alpha-1)}{\beta_e^2}\left(\frac{\lambda_e}{d_i}\right)\left(\frac{m_e}{m_i}\right)^\alpha\left(\frac{\delta B}{B} \right)^2\left(\frac{v}{v_A} \right)^{\alpha+2}\right].
\end{eqnarray}
The first term in the brackets corresponds to the classical Coulomb scattering, while the second term described the anomalous scattering by the whistlers. 

For further consideration, one needs to specify the parameters of the whistler turbulence: its spectral scaling $\alpha$, and the intensity of the turbulent fluctuations. As an example, we may perform a simple estimate by assuming that the field-perpendicular spectrum of the turbulence scales as $k_\perp^{-8/3}$ and its  anisotropy is $k_\| \propto (k_\perp)^{1/3}$. This is consistent with observations and numerical simulations~\cite[][]{howes_astrophysical_2006,cho2009,alexandrova09,kiyani09a,chen10b,chen12a,sahraoui13a,meyrand2013,Groselj2018,roytershteyn2019}. We then derive that the field-parallel spectrum scales as $\sim k_\|^{-6}$, and, therefore, $\alpha=6$. In addition, at the distance of $1$~AU, we may estimate 
$ d_i=10^7 \,{\rm cm}$, 
$ \lambda_e=10^{13}\, {\rm cm}$,
$ v_A=7\times 10^6\, {\rm cm/s}$,
$ v_{Te} = 2\times 10^8\, {\rm cm/s} \approx 10\, {\rm eV}$,
and $\beta_e=1$. For the intensity of the whistler fluctuations, we may follow observational results \cite[e.g.,][]{chen_b2013} and assume that the intensity of magnetic fluctuations is on order of $\left({\delta B}/{B_0}\right)^2\sim 10^{-2}$. In fact, it is believed that the magnetic fluctuations in the observations are dominated by the kinetic-Alfv\'en modes, with the whistlers contributing only a fraction of the fluctuation energy \cite[e.g.,][]{chen_b2013}, so this expression can serve as a rather conservative upper boundary.  We then obtain
\begin{eqnarray}
S  = \frac{4 \pi n(r) e^4 \Lambda  }{m_e^2 v^3}\left[1+ \left(\frac{\cal E}{{\cal E}_c} \right)^{4}\right],
\end{eqnarray}
where  ${\cal E}_c \approx 200\,{\rm eV}$.\footnote{This value, based of somewhat overestimated magnitude of the magnetic fluctuations $\delta B/B_0$, provides a lower boundary for the characteristic energy ${\cal E}_c$. As follows from formula~(\ref{Ec}) given below, weaker magnetic fluctuations would lead to a larger characteristic energy.} 

We see that the anomalous diffusion strongly depends on the energy. {For energies below the characteristic energy ${\cal E}_c$, the electron scattering is provided mostly by Coulomb collisions. As the energy increases above ${\cal E}_c$, scattering by whistlers rapidly becomes dominant.}   This result is consistent, for instance, with the fast solar wind observations in \cite[e.g.,][]{horaites2019} that found that the electron strahl is rather well described by the Coulomb theory, that is, it is not affected by anomalously strong scattering at relatively low energies, below $100 - 200$~eV. {Our results are also broadly consistent with the recent studies by \cite[][]{bercic2019} who noticed that the strahl angular broadening is a function of the electron energy and it starts to increase at energies exceeding several hundred~$eV$.}

From Eq.~(\ref{coll_op_eq}) one can see that the threshold energy scales as 
\begin{eqnarray}\label{Ec}
{\cal E}_c \propto \left[\left(\delta B\right)^{-2}B(r)^6 n(r)^{-3/2}\right]^{1/4}.
\end{eqnarray}
 According to observational estimates \cite[e.g.,][]{horbury1996,horbury2001,bruno2013}, the intensity of magnetic fluctuations measured at a given frequency in the Alfv\'enic inertial interval in a spacecraft frame (that, according to the Taylor hypothesis, corresponds to a given field-perpendicular scale), declines as $1/r^4$ with the heliospheric distance. Simultaneously, since the plasma density declines as $n(r)\propto 1/r^2$, the ion inertial scale increases with the distance as $d_i\propto r$. In order to estimate how the whistler component of the turbulence evolves with the distance, one needs to know the mechanism of turbulence generation, which is currently not well understood. One may, however, assume that the intensity of the whistler turbulence is proportional to the intensity of kinetic plasma turbulence at the scale $d_i$ (in fact, whistler turbulence may be generated at this scale \cite[e.g.,][]{horaites18b}). Since at scales smaller that $d_i$, the spectrum of observed magnetic fluctuation is $\sim k_\perp^{-8/3}$, their intensity at $d_i$ is  $\int_{1/d_i} k_\perp^{-8/3}dk_\perp \sim d_i^{5/3}$. When the  intensity of the Alfvenic fluctuations at a given scale decreases as $1/r^4$ while the transition scale to the kinetic regime increases as $d_i\sim r$, the intensity of magnetic fluctuations at the $d_i$ scale varies as $(\delta B/B)^2\propto r^{-4+5/3}=r^{-7/3}$. We therefore assume this behavior as the upper boundary for the whistler fluctuations.

In the inner heliosphere ($r\ll 1$~AU) the magnetic field strength scales approximately as $B(r)\propto 1/r^2$, therefore, the threshold ${\cal E}_c$ should vary approximately as ${\cal E}_c\propto r^{-5/3}$. In the outer heliosphere ($r\gg 1$~AU), the magnetic strength varies approximately as $B(r)\propto 1/r$, therefore, the energy threshold scales as ${\cal E}_c\propto r^{-1/6}$. The threshold is quite high in the inner heliosphere so it does not significantly affect the number of particles in the strahl. In the outer heliosphere, however, the threshold may be comparable to $T_0$, so its variations may significantly affect the number of strahl particles.  This may be broadly consistent with the observational results that the fraction of the electrons  forming the strahl decreases in the outer heliosphere.

\section{Conclusions}
{In this work, we have developed a kinetic theory of the electron strahl, which describes the global evolution of the strahl electrons and relates their velocity distribution function at the hot coronal region to that at larger heliospheric  distances. We have solved the drift-kinetic equation that traces the distribution function along the Parker-spiraled magnetic field lines. We have considered two pitch-angle scattering mechanisms that are believed to be relevant for the strahl broadening - Coulomb collisions (classical scattering) and scattering by plasma turbulence (anomalous scattering).} 
 
{The main prediction of our Coulomb theory is the strahl distribution function given by Eq.~(\ref{f_total}). We have found that this theory captures some essential physics of the strahl formation. In particular, the number of electrons forming the strahl, given by Eq.~(\ref{n_st}), and the angular width of the strahl, given by Eq.~(\ref{sin}), are in good qualitative and sometimes quantitative agreement with the available observations where there are reason to believe that anomalous scattering is not significant \cite[e.g.,][]{horaites18a,horaites18b,horaites2019}. }  

{When anomalous scattering mechanisms, e.g., pitch-angle diffusion caused by plasma turbulence, become important, the Coulomb theory is not applicable. To address such a situation, in addition to Coulomb collisions we have considered a quasi-linear diffusion provided by oblique whistler turbulence, as described by Eqs.~(\ref{C1}, \ref{coll_op_eq_2}). We have found that in this case, the angular broadening of the strahl becomes energy dependent. In particular, it alters the Coulomb theory at high energies. Whistler turbulence may, therefore, efficiently scatter and isotropize very energetic electrons possibly leading to formation of the electron halo. }

{According to our results, the anomalous scattering becomes significant when the electron kinetic energy exceeds certain characteristic energy ${\cal E}_c$. This energy threshold becomes lower at larger heliospheric distances, implying that the anomalous scattering mechanism becomes progressively more important with the distance. For a model spectral distribution of whistler turbulence we estimate that at 1AU, the anomalous scattering is not expected to be significant as compared to Coulomb collisions at the energies below $200$~eV,  but it becomes progressively more important at higher energies. These results are broadly consistent with the recent analytical and observational findings by~\cite{horaites2019,bercic2019}.}

\section*{Acknowledgements}
We are grateful to Christopher Chen for useful comments. The work of SB and KH was supported by the NSF under the grant no. NSF PHY-1707272 and by NASA under the grant no. NASA 80NSSC18K0646. SB was also partly supported by the DOE grant No. DE-SC0018266.





\begin{thebibliography}{}
\makeatletter
\relax
\def\mn@urlcharsother{\let\do\@makeother \do\$\do\&\do\#\do\^\do\_\do\%\do\~}
\def\mn@doi{\begingroup\mn@urlcharsother \@ifnextchar [ {\mn@doi@}
  {\mn@doi@[]}}
\def\mn@doi@[#1]#2{\def\@tempa{#1}\ifx\@tempa\@empty \href
  {http://dx.doi.org/#2} {doi:#2}\else \href {http://dx.doi.org/#2} {#1}\fi
  \endgroup}
\def\mn@eprint#1#2{\mn@eprint@#1:#2::\@nil}
\def\mn@eprint@arXiv#1{\href {http://arxiv.org/abs/#1} {{\tt arXiv:#1}}}
\def\mn@eprint@dblp#1{\href {http://dblp.uni-trier.de/rec/bibtex/#1.xml}
  {dblp:#1}}
\def\mn@eprint@#1:#2:#3:#4\@nil{\def\@tempa {#1}\def\@tempb {#2}\def\@tempc
  {#3}\ifx \@tempc \@empty \let \@tempc \@tempb \let \@tempb \@tempa \fi \ifx
  \@tempb \@empty \def\@tempb {arXiv}\fi \@ifundefined
  {mn@eprint@\@tempb}{\@tempb:\@tempc}{\expandafter \expandafter \csname
  mn@eprint@\@tempb\endcsname \expandafter{\@tempc}}}

\bibitem[\protect\citeauthoryear{{Alexandrova}, {Saur}, {Lacombe}, {Mangeney},
  {Mitchell}, {Schwartz}  \& {Robert}}{{Alexandrova}
  et~al.}{2009}]{alexandrova09}
{Alexandrova} O.,  {Saur} J.,  {Lacombe} C.,  {Mangeney} A.,  {Mitchell} J.,
  {Schwartz} S.~J.,   {Robert} P.,  2009, \mn@doi [Physical Review Letters]
  {10.1103/PhysRevLett.103.165003}, \href
  {http://adsabs.harvard.edu/abs/2009PhRvL.103p5003A} {103, 165003}

\bibitem[\protect\citeauthoryear{{Anderson}, {Skoug}, {Steinberg}  \&
  {McComas}}{{Anderson} et~al.}{2012}]{anderson12}
{Anderson} B.~R.,  {Skoug} R.~M.,  {Steinberg} J.~T.,   {McComas} D.~J.,  2012,
  \mn@doi [Journal of Geophysical Research (Space Physics)]
  {10.1029/2011JA017269}, \href
  {http://adsabs.harvard.edu/abs/2012JGRA..117.4107A} {117, A04107}

\bibitem[\protect\citeauthoryear{{Bale} et~al.,}{{Bale}
  et~al.}{2016}]{bale2016}
{Bale} S.~D.,  et~al., 2016, \mn@doi [Space Science Reviews]
  {10.1007/s11214-016-0244-5}, \href
  {http://adsabs.harvard.edu/abs/2016SSRv..204...49B} {204, 49}

\bibitem[\protect\citeauthoryear{{Ber{\v{c}}i{\v{c}}}, {Maksimovi{\'c}}, {},
  {Land i}  \& {Matteini}}{{Ber{\v{c}}i{\v{c}}} et~al.}{2019}]{bercic2019}
{Ber{\v{c}}i{\v{c}}} L.,  {Maksimovi{\'c}} {} M.,  {Land i} S.,   {Matteini}
  L.,  2019, \mn@doi [\mnras] {10.1093/mnras/stz1007}, \href
  {https://ui.adsabs.harvard.edu/abs/2019MNRAS.486.3404B} {486, 3404}

\bibitem[\protect\citeauthoryear{{Boldyrev} \& {Perez}}{{Boldyrev} \&
  {Perez}}{2012}]{boldyrev_p12}
{Boldyrev} S.,  {Perez} J.~C.,  2012, \mn@doi [The Astrophysical Journal
  Letters] {10.1088/2041-8205/758/2/L44}, \href
  {http://adsabs.harvard.edu/abs/2012ApJ...758L..44B} {758, L44}

\bibitem[\protect\citeauthoryear{{Bruno} \& {Carbone}}{{Bruno} \&
  {Carbone}}{2013}]{bruno2013}
{Bruno} R.,  {Carbone} V.,  2013, \mn@doi [Living Reviews in Solar Physics]
  {10.12942/lrsp-2013-2}, \href
  {http://adsabs.harvard.edu/abs/2013LRSP...10....2B} {10, 2}

\bibitem[\protect\citeauthoryear{{Chandran}, {Dennis}, {Quataert}  \&
  {Bale}}{{Chandran} et~al.}{2011}]{chandran11}
{Chandran} B.~D.~G.,  {Dennis} T.~J.,  {Quataert} E.,   {Bale} S.~D.,  2011,
  \mn@doi [\apj] {10.1088/0004-637X/743/2/197}, \href
  {http://adsabs.harvard.edu/abs/2011ApJ...743..197C} {743, 197}

\bibitem[\protect\citeauthoryear{{Chen}}{{Chen}}{2016}]{chen2016}
{Chen} C.~H.~K.,  2016, \mn@doi [Journal of Plasma Physics]
  {10.1017/S0022377816001124}, \href
  {http://adsabs.harvard.edu/abs/2016JPlPh..82f5302C} {82, 535820602}

\bibitem[\protect\citeauthoryear{Chen \& Boldyrev}{Chen \&
  Boldyrev}{2017}]{chen_boldyrev2017}
Chen C. H.~K.,  Boldyrev S.,  2017, \mn@doi [The Astrophysical Journal]
  {10.3847/1538-4357/aa74e0}, 842, 122

\bibitem[\protect\citeauthoryear{{Chen}, {Horbury}, {Schekochihin}, {Wicks},
  {Alexandrova}  \& {Mitchell}}{{Chen} et~al.}{2010}]{chen10b}
{Chen} C.~H.~K.,  {Horbury} T.~S.,  {Schekochihin} A.~A.,  {Wicks} R.~T.,
  {Alexandrova} O.,   {Mitchell} J.,  2010, \mn@doi [Physical Review Letters]
  {10.1103/PhysRevLett.104.255002}, \href
  {http://adsabs.harvard.edu/abs/2010PhRvL.104y5002C} {104, 255002}

\bibitem[\protect\citeauthoryear{{Chen}, {Salem}, {Bonnell}, {Mozer}  \&
  {Bale}}{{Chen} et~al.}{2012}]{chen12a}
{Chen} C.~H.~K.,  {Salem} C.~S.,  {Bonnell} J.~W.,  {Mozer} F.~S.,   {Bale}
  S.~D.,  2012, \mn@doi [Physical Review Letters]
  {10.1103/PhysRevLett.109.035001}, \href
  {http://adsabs.harvard.edu/abs/2012PhRvL.109c5001C} {109, 035001}

\bibitem[\protect\citeauthoryear{{Chen}, {Boldyrev}, {Xia}  \& {Perez}}{{Chen}
  et~al.}{2013}]{chen_b2013}
{Chen} C.~H.~K.,  {Boldyrev} S.,  {Xia} Q.,   {Perez} J.~C.,  2013, \mn@doi
  [Physical Review Letters] {10.1103/PhysRevLett.110.225002}, \href
  {http://adsabs.harvard.edu/abs/2013PhRvL.110v5002C} {110, 225002}

\bibitem[\protect\citeauthoryear{{Cho} \& {Lazarian}}{{Cho} \&
  {Lazarian}}{2009}]{cho2009}
{Cho} J.,  {Lazarian} A.,  2009, \mn@doi [\apj] {10.1088/0004-637X/701/1/236},
  \href {http://adsabs.harvard.edu/abs/2009ApJ...701..236C} {701, 236}

\bibitem[\protect\citeauthoryear{{Cranmer}, {Matthaeus}, {Breech}  \&
  {Kasper}}{{Cranmer} et~al.}{2009}]{cranmer2009}
{Cranmer} S.~R.,  {Matthaeus} W.~H.,  {Breech} B.~A.,   {Kasper} J.~C.,  2009,
  \mn@doi [\apj] {10.1088/0004-637X/702/2/1604}, \href
  {http://adsabs.harvard.edu/abs/2009ApJ...702.1604C} {702, 1604}

\bibitem[\protect\citeauthoryear{{Feldman}, {Asbridge}, {Bame}, {Montgomery}
  \& {Gary}}{{Feldman} et~al.}{1975}]{feldman75}
{Feldman} W.~C.,  {Asbridge} J.~R.,  {Bame} S.~J.,  {Montgomery} M.~D.,
  {Gary} S.~P.,  1975, \mn@doi [\jgr] {10.1029/JA080i031p04181}, \href
  {http://adsabs.harvard.edu/abs/1975JGR....80.4181F} {80, 4181}

\bibitem[\protect\citeauthoryear{{Forslund}}{{Forslund}}{1970}]{forslund70}
{Forslund} D.~W.,  1970, \mn@doi [\jgr] {10.1029/JA075i001p00017}, \href
  {http://adsabs.harvard.edu/abs/1970JGR....75...17F} {75, 17}

\bibitem[\protect\citeauthoryear{{Gary}, {Feldman}, {Forslund}  \&
  {Montgomery}}{{Gary} et~al.}{1975}]{gary75}
{Gary} S.~P.,  {Feldman} W.~C.,  {Forslund} D.~W.,   {Montgomery} M.~D.,  1975,
  \mn@doi [\jgr] {10.1029/JA080i031p04197}, \href
  {http://adsabs.harvard.edu/abs/1975JGR....80.4197G} {80, 4197}

\bibitem[\protect\citeauthoryear{{Gary}, {Scime}, {Phillips}  \&
  {Feldman}}{{Gary} et~al.}{1994}]{gary94}
{Gary} S.~P.,  {Scime} E.~E.,  {Phillips} J.~L.,   {Feldman} W.~C.,  1994,
  \mn@doi [\jgr] {10.1029/94JA02067}, \href
  {http://adsabs.harvard.edu/abs/1994JGR....9923391G} {99, 23391}

\bibitem[\protect\citeauthoryear{{Goldreich} \& {Sridhar}}{{Goldreich} \&
  {Sridhar}}{1995}]{goldreich_toward_1995}
{Goldreich} P.,  {Sridhar} S.,  1995, \mn@doi [The Astrophysical Journal]
  {10.1086/175121}, \href {http://adsabs.harvard.edu/abs/1995ApJ...438..763G}
  {438, 763}

\bibitem[\protect\citeauthoryear{{Gosling}, {Bame}, {Feldman}, {McComas},
  {Phillips}  \& {Goldstein}}{{Gosling} et~al.}{1993}]{gosling1993}
{Gosling} J.~T.,  {Bame} S.~J.,  {Feldman} W.~C.,  {McComas} D.~J.,  {Phillips}
  J.~L.,   {Goldstein} B.~E.,  1993, \mn@doi [\grl] {10.1029/93GL02489}, \href
  {http://adsabs.harvard.edu/abs/1993GeoRL..20.2335G} {20, 2335}

\bibitem[\protect\citeauthoryear{{Gosling}, {Skoug}  \& {Feldman}}{{Gosling}
  et~al.}{2001}]{gosling2001}
{Gosling} J.~T.,  {Skoug} R.~M.,   {Feldman} W.~C.,  2001, \mn@doi [\grl]
  {10.1029/2001GL013758}, \href
  {http://adsabs.harvard.edu/abs/2001GeoRL..28.4155G} {28, 4155}

\bibitem[\protect\citeauthoryear{{Graham} et~al.,}{{Graham}
  et~al.}{2017}]{graham17}
{Graham} G.~A.,  et~al., 2017, \mn@doi [Journal of Geophysical Research (Space
  Physics)] {10.1002/2016JA023656}, \href
  {http://adsabs.harvard.edu/abs/2017JGRA..122.3858G} {122, 3858}

\bibitem[\protect\citeauthoryear{{Graham}, {Rae}, {Owen}  \& {Walsh}}{{Graham}
  et~al.}{2018}]{graham18}
{Graham} G.~A.,  {Rae} I.~J.,  {Owen} C.~J.,   {Walsh} A.~P.,  2018, \mn@doi
  [\apj] {10.3847/1538-4357/aaaf1b}, \href
  {http://adsabs.harvard.edu/abs/2018ApJ...855...40G} {855, 40}

\bibitem[\protect\citeauthoryear{Gro{\v{s}}elj, Mallet, Loureiro  \&
  Jenko}{Gro{\v{s}}elj et~al.}{2018}]{Groselj2018}
Gro{\v{s}}elj D.,  Mallet A.,  Loureiro N.~F.,   Jenko F.,  2018, \mn@doi
  [Physical Review Letters] {10.1103/PhysRevLett.120.105101}, 120, 1

\bibitem[\protect\citeauthoryear{{Helander} \& {Sigmar}}{{Helander} \&
  {Sigmar}}{2002}]{helandersigmar02}
{Helander} P.,  {Sigmar} D.~J.,  2002, {Collisional transport in magnetized
  plasmas. Cambridge University Press (Cambridge monographs on plasma
  physics;~4)}

\bibitem[\protect\citeauthoryear{{Horaites}, {Boldyrev}, {Krasheninnikov},
  {Salem}, {Bale}  \& {Pulupa}}{{Horaites} et~al.}{2015}]{horaites15}
{Horaites} K.,  {Boldyrev} S.,  {Krasheninnikov} S.~I.,  {Salem} C.,  {Bale}
  S.~D.,   {Pulupa} M.,  2015, \mn@doi [Physical Review Letters]
  {10.1103/PhysRevLett.114.245003}, \href
  {http://adsabs.harvard.edu/abs/2015PhRvL.114x5003H} {114, 245003}

\bibitem[\protect\citeauthoryear{{Horaites}, {Boldyrev}, {Wilson}, {Vi{\~n}as}
  \& {Merka}}{{Horaites} et~al.}{2018a}]{horaites18a}
{Horaites} K.,  {Boldyrev} S.,  {Wilson} III L.~B.,  {Vi{\~n}as} A.~F.,
  {Merka} J.,  2018a, \mn@doi [\mnras] {10.1093/mnras/stx2555}, \href
  {http://adsabs.harvard.edu/abs/2018MNRAS.474..115H} {474, 115}

\bibitem[\protect\citeauthoryear{{Horaites}, {Astfalk}, {Boldyrev}  \&
  {Jenko}}{{Horaites} et~al.}{2018b}]{horaites18b}
{Horaites} K.,  {Astfalk} P.,  {Boldyrev} S.,   {Jenko} F.,  2018b, \mn@doi
  [\mnras] {10.1093/mnras/sty1808}, \href
  {http://adsabs.harvard.edu/abs/2018MNRAS.480.1499H} {480, 1499}

\bibitem[\protect\citeauthoryear{{Horaites}, {Boldyrev}  \&
  {Medvedev}}{{Horaites} et~al.}{2019}]{horaites2019}
{Horaites} K.,  {Boldyrev} S.,   {Medvedev} M.~V.,  2019, \mn@doi [\mnras]
  {10.1093/mnras/sty3504}, \href
  {http://adsabs.harvard.edu/abs/2019MNRAS.484.2474H} {484, 2474}

\bibitem[\protect\citeauthoryear{{Horbury} \& {Balogh}}{{Horbury} \&
  {Balogh}}{2001}]{horbury2001}
{Horbury} T.~S.,  {Balogh} A.,  2001, \mn@doi [\jgr] {10.1029/2000JA000108},
  \href {http://adsabs.harvard.edu/abs/2001JGR...10615929H} {106, 15929}

\bibitem[\protect\citeauthoryear{{Horbury}, {Balogh}, {Forsyth}  \&
  {Smith}}{{Horbury} et~al.}{1996}]{horbury1996}
{Horbury} T.~S.,  {Balogh} A.,  {Forsyth} R.~J.,   {Smith} E.~J.,  1996, \aap,
  \href {http://adsabs.harvard.edu/abs/1996A%26A...316..333H} {316, 333}

\bibitem[\protect\citeauthoryear{{Howes}, {Cowley}, {Dorland}, {Hammett},
  {Quataert}  \& {Schekochihin}}{{Howes}
  et~al.}{2006}]{howes_astrophysical_2006}
{Howes} G.~G.,  {Cowley} S.~C.,  {Dorland} W.,  {Hammett} G.~W.,  {Quataert}
  E.,   {Schekochihin} A.~A.,  2006, \mn@doi [The Astrophysical Journal]
  {10.1086/506172}, \href {http://adsabs.harvard.edu/abs/2006ApJ...651..590H}
  {651, 590}

\bibitem[\protect\citeauthoryear{{Kajdi{\v c}}, {Alexandrova}, {Maksimovic},
  {Lacombe}  \& {Fazakerley}}{{Kajdi{\v c}} et~al.}{2016}]{kajdic16}
{Kajdi{\v c}} P.,  {Alexandrova} O.,  {Maksimovic} M.,  {Lacombe} C.,
  {Fazakerley} A.~N.,  2016, \mn@doi [\apj] {10.3847/1538-4357/833/2/172},
  \href {http://adsabs.harvard.edu/abs/2016ApJ...833..172K} {833, 172}

\bibitem[\protect\citeauthoryear{{Kiyani}, {Chapman}, {Khotyaintsev}, {Dunlop}
  \& {Sahraoui}}{{Kiyani} et~al.}{2009}]{kiyani09a}
{Kiyani} K.~H.,  {Chapman} S.~C.,  {Khotyaintsev} Y.~V.,  {Dunlop} M.~W.,
  {Sahraoui} F.,  2009, \mn@doi [Physical Review Letters]
  {10.1103/PhysRevLett.103.075006}, \href
  {http://adsabs.harvard.edu/abs/2009PhRvL.103g5006K} {103, 075006}

\bibitem[\protect\citeauthoryear{{K{\"o}hnlein}}{{K{\"o}hnlein}}{1996}]{koehnlein96}
{K{\"o}hnlein} W.,  1996, \mn@doi [\solphys] {10.1007/BF00153841}, \href
  {http://adsabs.harvard.edu/abs/1996SoPh..169..209K} {169, 209}

\bibitem[\protect\citeauthoryear{{Kulsrud}}{{Kulsrud}}{2005}]{kulsrud2005}
{Kulsrud} R.~M.,  2005, {Plasma physics for astrophysics. Princeton University
  Press (Princeton series in astrophysics)}

\bibitem[\protect\citeauthoryear{{Lacombe}, {Alexandrova}, {Matteini},
  {Santol{\'{\i}}k}, {Cornilleau-Wehrlin}, {Mangeney}, {de Conchy}  \&
  {Maksimovic}}{{Lacombe} et~al.}{2014}]{lacombe14}
{Lacombe} C.,  {Alexandrova} O.,  {Matteini} L.,  {Santol{\'{\i}}k} O.,
  {Cornilleau-Wehrlin} N.,  {Mangeney} A.,  {de Conchy} Y.,   {Maksimovic} M.,
  2014, \mn@doi [\apj] {10.1088/0004-637X/796/1/5}, \href
  {http://adsabs.harvard.edu/abs/2014ApJ...796....5L} {796, 5}

\bibitem[\protect\citeauthoryear{{Meyrand} \& {Galtier}}{{Meyrand} \&
  {Galtier}}{2013}]{meyrand2013}
{Meyrand} R.,  {Galtier} S.,  2013, \mn@doi [Physical Review Letters]
  {10.1103/PhysRevLett.111.264501}, \href
  {http://adsabs.harvard.edu/abs/2013PhRvL.111z4501M} {111, 264501}

\bibitem[\protect\citeauthoryear{{Narita} et~al.,}{{Narita}
  et~al.}{2016}]{narita16}
{Narita} Y.,  et~al., 2016, \mn@doi [\apjl] {10.3847/2041-8205/827/1/L8}, \href
  {http://adsabs.harvard.edu/abs/2016ApJ...827L...8N} {827, L8}

\bibitem[\protect\citeauthoryear{{Ogilvie}, {Fitzenreiter}  \&
  {Desch}}{{Ogilvie} et~al.}{2000}]{ogilvie00}
{Ogilvie} K.~W.,  {Fitzenreiter} R.,   {Desch} M.,  2000, \mn@doi [\jgr]
  {10.1029/2000JA000131}, \href
  {http://adsabs.harvard.edu/abs/2000JGR...10527277O} {105, 27277}

\bibitem[\protect\citeauthoryear{{Pagel}, {Gary}, {de Koning}, {Skoug}  \&
  {Steinberg}}{{Pagel} et~al.}{2007}]{pagel07}
{Pagel} C.,  {Gary} S.~P.,  {de Koning} C.~A.,  {Skoug} R.~M.,   {Steinberg}
  J.~T.,  2007, \mn@doi [Journal of Geophysical Research (Space Physics)]
  {10.1029/2006JA011967}, \href
  {http://adsabs.harvard.edu/abs/2007JGRA..112.4103P} {112, A04103}

\bibitem[\protect\citeauthoryear{{Pierrard}, {Maksimovic}  \&
  {Lemaire}}{{Pierrard} et~al.}{2001}]{pierrard01}
{Pierrard} V.,  {Maksimovic} M.,   {Lemaire} J.,  2001, \mn@doi [\jgr]
  {10.1029/2001JA900133}, \href
  {http://adsabs.harvard.edu/abs/2001JGR...10629305P} {106, 29305}

\bibitem[\protect\citeauthoryear{{Pierrard}, {Lazar}  \&
  {Schlickeiser}}{{Pierrard} et~al.}{2011}]{pierrard2011}
{Pierrard} V.,  {Lazar} M.,   {Schlickeiser} R.,  2011, \mn@doi [\solphys]
  {10.1007/s11207-010-9700-7}, \href
  {http://adsabs.harvard.edu/abs/2011SoPh..269..421P} {269, 421}

\bibitem[\protect\citeauthoryear{{Pierrard}, {Lazar}, {Poedts}, {{\v
  S}tver{\'a}k}, {Maksimovic}  \& {Tr{\'a}vn{\'{\i}}{\v c}ek}}{{Pierrard}
  et~al.}{2016}]{pierrard2016}
{Pierrard} V.,  {Lazar} M.,  {Poedts} S.,  {{\v S}tver{\'a}k} {\v S}.,
  {Maksimovic} M.,   {Tr{\'a}vn{\'{\i}}{\v c}ek} P.~M.,  2016, \mn@doi
  [\solphys] {10.1007/s11207-016-0961-7}, \href
  {http://adsabs.harvard.edu/abs/2016SoPh..291.2165P} {291, 2165}

\bibitem[\protect\citeauthoryear{{Pilipp}, {Muehlhaeuser}, {Miggenrieder},
  {Montgomery}  \& {Rosenbauer}}{{Pilipp} et~al.}{1987}]{pilipp87}
{Pilipp} W.~G.,  {Muehlhaeuser} K.-H.,  {Miggenrieder} H.,  {Montgomery} M.~D.,
    {Rosenbauer} H.,  1987, \mn@doi [\jgr] {10.1029/JA092iA02p01075}, \href
  {http://adsabs.harvard.edu/abs/1987JGR....92.1075P} {92, 1075}

\bibitem[\protect\citeauthoryear{{Roytershteyn}, {Boldyrev}, {Delzanno},
  {Chen}, {Gro{\v s}elj}  \& {Loureiro}}{{Roytershteyn}
  et~al.}{2019}]{roytershteyn2019}
{Roytershteyn} V.,  {Boldyrev} S.,  {Delzanno} G.~L.,  {Chen} C.~H.~K.,
  {Gro{\v s}elj} D.,   {Loureiro} N.~F.,  2019, \mn@doi [The Astrophysical
  Journal] {10.3847/1538-4357/aaf288}, \href
  {http://adsabs.harvard.edu/abs/2019ApJ...870..103R} {870, 103}

\bibitem[\protect\citeauthoryear{{Sahraoui}, {Huang}, {Belmont}, {Goldstein},
  {R{\'e}tino}, {Robert}  \& {De Patoul}}{{Sahraoui}
  et~al.}{2013}]{sahraoui13a}
{Sahraoui} F.,  {Huang} S.~Y.,  {Belmont} G.,  {Goldstein} M.~L.,  {R{\'e}tino}
  A.,  {Robert} P.,   {De Patoul} J.,  2013, \mn@doi [The Astrophysical
  Journal] {10.1088/0004-637X/777/1/15}, \href
  {http://adsabs.harvard.edu/abs/2013ApJ...777...15S} {777, 15}

\bibitem[\protect\citeauthoryear{{Saito} \& {Gary}}{{Saito} \&
  {Gary}}{2007}]{saitogary07}
{Saito} S.,  {Gary} S.~P.,  2007, \mn@doi [\grl] {10.1029/2006GL028173}, \href
  {http://adsabs.harvard.edu/abs/2007GeoRL..34.1102S} {34, L01102}

\bibitem[\protect\citeauthoryear{{Scudder} \& {Olbert}}{{Scudder} \&
  {Olbert}}{1979}]{scudderolbert79}
{Scudder} J.~D.,  {Olbert} S.,  1979, \mn@doi [\jgr] {10.1029/JA084iA11p06603},
  \href {http://adsabs.harvard.edu/abs/1979JGR....84.6603S} {84, 6603}

\bibitem[\protect\citeauthoryear{{Stansby}, {Horbury}, {Chen}  \&
  {Matteini}}{{Stansby} et~al.}{2016}]{stansby16}
{Stansby} D.,  {Horbury} T.~S.,  {Chen} C.~H.~K.,   {Matteini} L.,  2016,
  \mn@doi [\apjl] {10.3847/2041-8205/829/1/L16}, \href
  {http://adsabs.harvard.edu/abs/2016ApJ...829L..16S} {829, L16}

\bibitem[\protect\citeauthoryear{{Stix}}{{Stix}}{1992}]{stix1992}
{Stix} T.~H.,  1992, {Waves in plasmas. New York : American Institute of
  Physics}

\bibitem[\protect\citeauthoryear{{Tang}, {Zank}  \& {Kolobov}}{{Tang}
  et~al.}{2018}]{tang2018}
{Tang} B.,  {Zank} G.~P.,   {Kolobov} V.,  2018, in Journal of Physics
  Conference Series. p. 012025, \mn@doi{10.1088/1742-6596/1100/1/012025}

\bibitem[\protect\citeauthoryear{{TenBarge} \& {Howes}}{{TenBarge} \&
  {Howes}}{2012}]{tenbarge2012}
{TenBarge} J.~M.,  {Howes} G.~G.,  2012, \mn@doi [Physics of Plasmas]
  {10.1063/1.3693974}, \href
  {http://adsabs.harvard.edu/abs/2012PhPl...19e5901T} {19, 055901}

\bibitem[\protect\citeauthoryear{{Verscharen}, {Klein}  \&
  {Maruca}}{{Verscharen} et~al.}{2019}]{verscharen2019}
{Verscharen} D.,  {Klein} K.~G.,   {Maruca} B.~A.,  2019, arXiv e-prints, \href
  {http://adsabs.harvard.edu/abs/2019arXiv190203448V} {}

\bibitem[\protect\citeauthoryear{{Vocks} \& {Mann}}{{Vocks} \&
  {Mann}}{2003}]{vocks03}
{Vocks} C.,  {Mann} G.,  2003, \mn@doi [\apj] {10.1086/376682}, \href
  {http://adsabs.harvard.edu/abs/2003ApJ...593.1134V} {593, 1134}

\bibitem[\protect\citeauthoryear{{Vocks}, {Salem}, {Lin}  \& {Mann}}{{Vocks}
  et~al.}{2005}]{vocks05}
{Vocks} C.,  {Salem} C.,  {Lin} R.~P.,   {Mann} G.,  2005, \mn@doi [\apj]
  {10.1086/430119}, \href {http://adsabs.harvard.edu/abs/2005ApJ...627..540V}
  {627, 540}

\bibitem[\protect\citeauthoryear{{Wilson} et~al.,}{{Wilson}
  et~al.}{2013}]{wilson13}
{Wilson} L.~B.,  et~al., 2013, \mn@doi [Journal of Geophysical Research (Space
  Physics)] {10.1029/2012JA018167}, \href
  {http://adsabs.harvard.edu/abs/2013JGRA..118....5W} {118, 5}

\bibitem[\protect\citeauthoryear{{{\v S}tver{\'a}k}, {Maksimovic},
  {Tr{\'a}vn{\'{\i}}{\v c}ek}, {Marsch}, {Fazakerley}  \& {Scime}}{{{\v
  S}tver{\'a}k} et~al.}{2009}]{stverak09}
{{\v S}tver{\'a}k} {\v S}.,  {Maksimovic} M.,  {Tr{\'a}vn{\'{\i}}{\v c}ek}
  P.~M.,  {Marsch} E.,  {Fazakerley} A.~N.,   {Scime} E.~E.,  2009, \mn@doi
  [Journal of Geophysical Research (Space Physics)] {10.1029/2008JA013883},
  \href {http://adsabs.harvard.edu/abs/2009JGRA..11405104S} {114, 5104}

\bibitem[\protect\citeauthoryear{{{\v S}tver{\'a}k}, {Tr{\'a}vn{\'{\i}}{\v
  c}ek}  \& {Hellinger}}{{{\v S}tver{\'a}k} et~al.}{2015}]{stverak15}
{{\v S}tver{\'a}k} {\v S}.,  {Tr{\'a}vn{\'{\i}}{\v c}ek} P.~M.,   {Hellinger}
  P.,  2015, \mn@doi [Journal of Geophysical Research (Space Physics)]
  {10.1002/2015JA021368}, \href
  {http://adsabs.harvard.edu/abs/2015JGRA..120.8177A} {120, 8177}

\makeatother
\end{thebibliography}





\bsp	
\label{lastpage}
\end{document}